\documentclass[a4paper,12pt]{article}
\usepackage{t1enc}
\usepackage[english]{babel}
\usepackage{amsmath, amssymb, amsthm, amsfonts}
\usepackage{multirow, slashbox, comment, hyperref}
\theoremstyle{definition}
\newtheorem{defn}{Definition}
\newtheorem*{rmk}{Remark}
\newtheorem*{rmks}{Remarks}
\newcommand{\complex}{\mathbb{C}}
\newcommand{\real}{\mathbb{R}}
\newcommand{\IN}{\mathbb{N}}
\newcommand{\IZ}{\mathbb{Z}}
\newcommand{\IR}{\mathbb{R}}
\newcommand{\IC}{\mathbb{C}}
\newcommand{\II}{\mathbb{I}}
\newcommand{\HH}{\mathcal{H}}
\newcommand{\BB}{\mathcal{B}}
\newcommand{\DD}{\mathcal{D}}
\newcommand{\id}{\text{id}}
\newcommand{\dom}{\text{dom}}

\def\ts{\otimes}

\begin{document}
\title{Product of real spectral triples}
\author{Ludwik D\k{a}browski, Giacomo Dossena\\ SISSA, Via Bonomea 265, 34136 Trieste, Italy}

\date{\empty}

\maketitle

\begin{abstract}
We construct the product of real spectral triples of arbitrary finite dimension (and arbitrary parity) taking into account the fact that in the even case there are two possible real structures, in the odd case there are two inequivalent representations of the gamma matrices (Clifford algebra), and in the even-even case there are two natural candidates for the Dirac operator of the product triple.
\end{abstract}

\section{Introduction}

In noncommutative geometry \`a la Connes a Riemannian spin manifold is encoded in terms of a spectral triple which satisfies seven additional properties \cite{C96}, \cite{G-BVF}, \cite{C08}.
These further properties have been formulated for noncommutative spectral triples \cite{C96}.
They are satisfied e.g by the noncommutative torus, and part of them holds for sundry quantum groups.

The composition of not necessarily commutative spectral triples corresponding to the Cartesian product of manifolds is of relevance 
for construction of a would-be tensor category, but also bears interest for some applications in theoretical physics.
For instance the almost commutative spectral triple corresponding to the standard model of particle physics \cite{C07} is a tensor product of a canonical commutative spectral triple with a finite dimensional noncommutative one.
Moreover the tensor product with a spectral triple of complex dimension is used in the Connes-Marcolli treatment \cite{C08} of dimensional renormalization of quantum fields.

In this paper we study the behaviour under tensor product of one of the additional properties, namely the reality axiom. 
This axiom is much more important in the noncommutative case as it is there employed in the formulation of few of the other axioms.

We carefully analyse all the possibilities and note that in even dimensions there are always two real structure operators $J$,
that differ by multiplication by the grading operator.
None of them should be preferred as they are perfectly on the same footing.
This leads to a richer table of their possible tensor products, which we study systematically, completing the results of \cite{Vanhecke} (see also \cite{Sitarz}) obtained for the even-even case and the even-odd (or odd-even) case. 
We construct also the tensor product of two odd real spectral triples (that requires a doubling of the tensor product of the Hilbert spaces).
When dealing with odd spectral triples we are careful about the two inequivalent representations of gamma matrices (Clifford algebra).

Composing two even-dimensional Dirac operators we consider two choices, which differ by using the chirality operator (grading) either of the first or of the second space. The two operators thus obtained are unitarily equivalent if no other requirements are imposed, but this is no longer the case when boundaries are present (\cite{CC}).
Moreover, the first expression is relevant for the composition of an even dimensional space with an odd dimensional one, while the second expression is relevant for the composition of an odd dimensional space with an even dimensional one.

For concreteness, we provide the explicit formulae for the eigenvalues and eigenvectors in terms of those of the individual components.
Furthermore, we also analyse few of the additional conditions (axioms) like dimension, regularity, first order and orientation.

\section{Commutative real spectral triple}\label{commutative}

In the following, the symbol $\IZ_+$ will denote the set of strictly positive integers and the symbol $\IN$ will denote the set of non-negative integers.

\subsection{Gamma matrices}
For each $n=\IZ_+$ consider the irreducible (complex) representations of the Clifford algebra $\mathcal{C}(\IR^n)$ of Euclidean space $\IR^n$ with negative-definite metric. For even (resp. odd) $n$, let us denote by $\Gamma_{(n)}$ (resp. $\Gamma_{(n,+)}$ and $\Gamma_{(n,-)}$) a possible choice of sets of complex matrices generating the only irreducible representation (respectively the only two irreducible representations) of $\mathcal{C}(\IR^n)$, given by:
\begin{equation}\label{rep}
\begin{aligned}
\Gamma_{(1,+)}&=\{\gamma^1_{(1,+)}\}\\
\Gamma_{(1,-)}&=\{\gamma^1_{(1,-)}\}\\
\Gamma_{(2m)}&=\{\gamma^1_{(2m)},\dots,\gamma^{2m}_{(2m)}\}\\
\Gamma_{(2m+1,+)}&=\{\gamma^1_{(2m+1,+)},\dots,\gamma^{2m}_{(2m+1,+)},\gamma^{2m+1}_{(2m+1,+)}\}\\
\Gamma_{(2m+1,-)}&=\{\gamma^1_{(2m+1,-)},\dots,\gamma^{2m}_{(2m+1,-)},\gamma^{2m+1}_{(2m+1,-)}\}\ ,
\end{aligned}
\end{equation}
where $\gamma^1_{(1,\pm)}=\pm i$ and for $n=2m$ ($m\in\IZ_+$), $j=1,\dots, m$, each $\gamma^\mu_{(n)}$ ($\mu=1,\dots,n$) is a $2^m\times 2^m$ complex matrix given by
\begin{equation}
\label{gammas-e}
\begin{aligned}
\gamma^j_{(n)} &=  i \underbrace{\sigma_3 \ts\cdots   \ts \sigma_3}_{m-j} \ts\, \sigma_1\ts \underbrace{\mathbf{1}\ts \cdots \ts \mathbf{1}}_{j-1}\ ,\\
\gamma^{m+j}_{(n)} &= i \underbrace{\sigma_3 \ts\cdots   \ts \sigma_3}_{m-j}
\ts\, \sigma_2\ts \underbrace{\mathbf{1}\ts \cdots \ts \mathbf{1}}_{j-1}\ ,
\end{aligned}
\end{equation}
while for $n=2m+1$ ($m\in\IZ_+$), $k=1,\dots,n-1$ we put
\begin{equation}
\label{gammas-o}
\begin{aligned}
\gamma^k_{(n,\pm)}&=\gamma^k_{(n-1)}\ ,\\
\gamma^{n}_{(n,\pm)} &= \pm i \underbrace{\sigma_3 \ts \cdots \ts \sigma_3}_{m}\ ,
\end{aligned}
\end{equation}
where
\begin{equation}\label{pauli}
\mathbf{1}=\left(
\begin{array}{cc}
1 & 0 \\
0 & 1 
\end{array}\right),\quad
\sigma_1=\left(
\begin{array}{cc}
0 & 1 \\
1 & 0 
\end{array}\right),\quad
\sigma_2=\left(
\begin{array}{cc}
0 & -i \\
i & 0 
\end{array}\right),\quad
\sigma_3=\left(
\begin{array}{cc}
1 & 0 \\
0 & -1 
\end{array}\right).
\end{equation}
Indeed, the gamma matrices defined above anti-commute with each other and square to $-1$. Note that they are anti-hermitean. Moreover we have chosen them so that for a given $n$ the first $m$ are imaginary, the next $m$ are real and $\gamma^{2m+1}_{(2m+1,\pm)}$ is imaginary.
The properties discussed in the sequel do not depend on this choice up to unitary equivalence of matrices.

Note also that
\begin{equation}\label{gamma_products}
\gamma^{2m+1}_{(2m+1,\pm)} = \pm i (i)^{(m\bmod 2)}\gamma^{1}_{(2m)}\gamma^{2}_{(2m)}\dots \gamma^{2m}_{(2m)}
\end{equation}
For even $n$ we define the grading operator $\chi_{(n)}:=\underbrace{\sigma_3 \ts \cdots \ts \sigma_3}_{n/2}$. For odd $n$ we define also $\chi_{(n, \pm)}:=\pm\mathbf{1}$. (In the sequel we shall often omit the lower indices to simplify notation). Note that, for any $n\in\IZ_+$,
\begin{equation}
\chi=\alpha_n \gamma^1\cdots \gamma^n\, ,
\end{equation}
where $\alpha_n=1,-i,i,1$ if $n=0,1,2,3\bmod 4$ respectively.

\subsection{Dirac operator}\label{Dirac_op}
The (free) Dirac operator on $\IR^n$ is given by the formula
\begin{equation}\label{Dop}
D = \sum_{\mu=1}^n \gamma^{\mu} \frac{\partial}{\partial x^{\mu}}.
\end{equation}
where the gammas are as above.
Note that, for even $n$, $$D\chi + \chi D =0.$$ 
Note also that, for odd $n$, changing the representation $\Gamma_{(2m+1,+)}$ to $\Gamma_{(2m+1,-)}$ is equivalent to changing the orientation of the manifold.

The `minimal coupling' interaction with gauge fields, notably the electromagnetic potential $A_\mu$, amounts to the substitution of the usual derivatives by the covariant ones,
$$\nabla_\mu = \frac{\partial}{\partial x^{\mu}} + i e A_\mu\, ,$$
where $e$ is the charge.

As well known, the Dirac operator can be defined for a flat metric of arbitrary signature, and generalized to (pseudo) Riemannian spin manifolds with the help of covariant derivative given by the Levi-Civita (spin) connection.
The elliptic or Riemannian case is extremely important and well studied in mathematics. In theoretical physics the Lorentzian case describes the evolution of spinor fields (fermions), and is also useful in connection with general relativity, modern versions of Kaluza-Klein theories, and (super) string theory. Recently, A. Connes made the Dirac operator a fundamental ingredient of a `spectral triple' and of the notion of noncommutative (spin) manifold.

\subsection{Charge conjugation}
In physics, the charge conjugation $J$ of spinors exchanges the Dirac operators corresponding to charge $e$ and $-e$, keeping invariant the other physical quantities. However, we allow a possibility that $J$ either commutes or anticommutes with the `neutral' $D$ given by (\ref{Dop}). We shall indicate by a subscript $\pm$ these two options, which amount to demanding that

\begin{equation}
\label{J}
J_\pm \gamma^{\mu} \left(\frac{\partial}{\partial x^{\mu}} + i e A_\mu\right)
= \pm \gamma^{\mu} \left(\frac{\partial}{\partial x^{\mu}} - i e A_\mu\right) J_\pm\ .
\end{equation}

The operators $J_\pm$ have to be $\IC$-antilinear, given by a composition of the complex conjugation with a constant matrix $C_\pm$, satisfying $C_\pm{\overline{\gamma}^\mu} = \pm  \gamma^\mu C_\pm$. Hence, $C_+$ should anti-commute with $\gamma^\mu$ for $\mu\leq m$ which are imaginary and commute with $\gamma^\mu$ for $m< \mu \leq 2m$ which are real. By the uniqueness and anti-commutativity of gamma matrices, such $C_+$ is proportional to $\gamma^1  \gamma^2 \cdots \gamma^m$ if $m$ is even and to $\gamma^{m+1}  \gamma^{m+2} \cdots \gamma^{2m}$ if $m$ is odd. It is just the other way for $C_-$.

For $\Gamma_{(2m)}$, this fixes the two solutions $J_\pm$ (up to a scalar multiple). Moreover, $J_-$ is obtained by multiplying $J_+$ with $\gamma^1\cdots\gamma^{2m}$ (up to a scalar multiple).

For $\Gamma_{(2m+1,+)}$ and for $\Gamma_{(2m+1,-)}$ we have to consider in addition the matrix $\gamma^{2m+1}_{(2m+1,\pm)}$, which is imaginary. Then the above (anti)-commutativity requirement selects $J_+$ as the only solution if $m=1, 3$ mod $4$ (i.e. $n=3, 7$ mod $8$)
and $J_-$ if $m=0, 2$ mod $4$ (i.e. $n=1, 5$ mod $8$).

It can be checked that, with respect to the standard Hermitean scalar product $<~,~>$ on ${\IC}^{2^m}$, the charge conjugation is a $\IC$-antilinear isometry, that is $JJ^\dagger=1=J^\dagger J$, where the adjoint of a $\IC$-antilinear operator is defined by $<\phi, J^\dagger \psi>=<\psi,J\phi>$. This reduces the ambiguity of $J_\pm$ to be a scalar of modulus 1.

The commutation relation of $J_\pm$ with $D$ is by construction $D J_\pm = \varepsilon^\prime J_\pm  D$, with $\varepsilon^\prime = +1$ 
for $J_+$ and $\varepsilon^\prime = -1$ for $J_-$. Next, the commutation relation with $\chi$ (if $n$ is even) is governed by $\varepsilon^{\prime\prime} = i^{n} = (-1)^{n/2}$. A straightforward computation gives $(J_\pm)^2 = \varepsilon \II$, where $\varepsilon$ (together with $\varepsilon^\prime, \varepsilon^{\prime\prime}$) is given by table \ref{ko-dim} below.

\begin{table}[!ht]\caption{Connes' selection in \cite{C08} is marked by $\bullet$}\label{ko-dim}
\centering
\begin{tabular}{|c||c|c|c|c||c|c|c|c||c|c|c|c|}
\hline
$n$	&	0	&	2	&	4	&	6	&	0	&	2	&	4	&	6		&	1	&	3	&	5	&	7\\
\hline
$\epsilon$	&	$+$	&	$-$	&	$-$	&	$+$	&	$+$	&	$+$	&	$-$	&	$-$	&	$+$	&	$-$	&	$-$	&	$+$\\
$\epsilon'$	&	$+$	&	$+$	&	$+$	&	$+$	&	$-$	&	$-$	&	$-$	&	$-$	&	$-$	&	$+$	&	$-$	&	$+$\\
$\epsilon''$	&	$+$	&	$-$	&	$+$	&	$-$	&	$+$	&	$-$	&	$+$	&	$-$	&	&	&	&\\
\hline
&	$\bullet$	&$\bullet$&$\bullet$&$\bullet$&	&	&	&	&$\bullet$&$\bullet$&$\bullet$&$\bullet$\\
\hline
\end{tabular}
\end{table}

Notice that altogether there are twelve different possibilities, which can be labeled by the so-called KO-dimension $n\in\IZ_8$ with the additional index $\epsilon'$ if $n$ is even (so for example the case $(\epsilon, \epsilon',\epsilon'')=(+,-,-)$ is labelled by $2_-$).
We find it notationally convenient to place this additional index also in the case of odd $n$, though it is redundant there. (For pseudoeuclidean spaces the periodicity modulo 8  holds for the signature $p-q$ of the metric).

The geometrical significance of the charge conjugation $J_\pm$ is that it governs the reduction of a spin$^c$ structure to a spin structure
(the Lie algebra $spin(n)$ is generated by $\gamma_{\mu}  \gamma_{\nu}$ with $\mu<\nu$, which commute with $J$ and so are invariant under $Ad_J$, while $spin^c(n)$ is generated by $spin(n)$ and one more matrix $i \II$, which anti-commutes with $J$).

The operator in (\ref{Dop}) is a first-order partial differential operator with matrix coefficients. It acts on $\mathcal{C}^\infty(\IR^n,\complex^{2^m})$. After completion to $L^2(\IR^n,\complex^{2^m})$, $D$ becomes an unbounded self-adjoint operator. The $*$-algebra of smooth complex-valued functions on $\IR^n$ (with pointwise operations) is represented on $L^2(\IR^n,\complex^{2^m})$ as multiplication operators.

\subsection{Dirac operator on Cartesian product}

Let $n=n_1+n_2$ with $n_1, n_2\in\IZ_+$. It is straightforward to see that the Dirac operator \eqref{Dop} on $\IR^n$ decomposes 
(up to unitary equivalences of matrices and a suitable renaming of coordinates) into Dirac operators $D_1$ on $\IR^{n_1}$ and $D_2$ on $\IR^{n_2}$ as follows ($\mathbf{1}$ and $\chi$ denote the relevant identity and grading matrices):

\begin{itemize}
\item if $n_1=2m_1$ and $n_2=2m_2+1$ ($m_1\in\IZ_+$, $m_2\in\IN$)

\begin{equation}\label{Deo}
D = 
\sum_{\mu =1}^{n_1} \gamma^{\mu}_{(n_1)}\!\ts\mathbf{1}\,
\frac{\partial}{\partial x^{\mu}}
+ 
\sum_{\nu =1}^{n_2} \chi_{(n_1)}\!\ts \gamma^{\nu}_{(n_2,\pm)}\, 
\frac{\partial}{\partial x^{n_1+\nu}}\, ,
\end{equation}
(using (\ref{gamma_products}) it is not difficult to see that the tensor product gamma matrices appearing in (\ref{Deo}) belong to the representation $\Gamma_{(2m_1+2m_2+1,\pm)}$, with the index $\pm$ identical to the one of $\gamma^\nu_{(n_2,\pm)}$, belonging to $\Gamma_{(2m_2+1,\pm)}$; in other words, the $\pm$-type of the irreducible representation is preserved);
\item if $n_1=2m_1+1$ and $n_2=2m_2$ ($m_1\in\IN$, $m_2\in\IZ_+$)

\begin{equation}\label{Doe}
D = 
\sum_{\mu =1}^{n_1} \gamma^{\mu}_{(n_1,\pm)}\!\ts\chi_{(n_2)}\,
\frac{\partial}{\partial x^{n_2+\mu}}
+ 
\sum_{\nu =1}^{n_2} \mathbf{1}\!\ts\gamma^{\nu}_{(n_2)}\, 
\frac{\partial}{\partial x^{\nu}}\, ,
\end{equation}
(again the $\pm$-type of the irreducible representation is preserved);

\item if both $n_1=2m_1$ and $n_2=2m_2$ ($m_1, m_2\in\IZ_+$) are even, both formulae 
\ref{Deo} and \ref{Doe} hold and are related by a unitary matrix;
\item if both $n_1=2m_1+1$ and $n_2=2m_2+1$ ($m_1, m_2\in\IN$) are odd then
\begin{equation}\label{Doo}
D = 
\sum_{\mu =1}^{n_1} \gamma^{\mu}_{(n_1)}\!\ts\mathbf{1}\ts \sigma_1\,
\frac{\partial}{\partial x^{\mu}}
+ 
\sum_{\nu =1}^{n_2} \mathbf{1}\!\ts \gamma^{\nu}_{(n_2)}\ts \sigma_2\, 
\frac{\partial}{\partial x^{n_1+\nu}}\, .
\end{equation}
Moreover, we can take $ \chi:=1\ts 1\ts \sigma_3$ as grading.
\end{itemize}

\section{Definition of a real spectral triple}

The classical setting presented in section \ref{commutative} was generalized by Connes to the noncommutative case, which we now recall and supplement by keeping all the twelve possibilities for the reality structure. We recall from \cite{C08}
\begin{defn}
A spectral triple $(A,\HH,D)$ is given by an involutive unital algebra $A$ (over $\real$ or $\complex$) faithfully represented as bounded operators on a complex separable Hilbert space $\HH$ and by a self-adjoint operator $D$ with compact resolvent such that for each $a\in A$ the commutator\footnote{We assume $a\dom D\subset\dom D$ for each $a\in A$, so that $[D,a]$ is defined on $\dom D$.} $[D, a]$ has bounded extension.

A spectral triple is called even if the Hilbert space $\HH$ is endowed with a nontrivial $\IZ_2$-grading $\chi$ which\footnote{By definition $\chi$ is a self-adjoint unitary such that $\chi^2=\id_\HH$ and $\chi\neq\pm\id_\HH$. The Hilbert space $\HH$ can then be split into its eigenspaces $\HH=\HH_+\oplus\HH_-$; by requesting $[\chi,a]=0$ for each $a\in A$ this splitting is invariant under the action of $A$ on $\HH$.} commutes with any $a\in A$ and anticommutes with $D$. Otherwise it is called odd.
\end{defn}

The following definition is a modification of Definition 1.124 in \cite{C08} in order to cover all the twelve possibilities as discussed in the previous section.

\begin{defn}
 A real structure of $KO$-dimension $n\in\IZ_8$ on a spectral triple $(A,\HH,D)$ is an antilinear isometry $J : \HH\to\HH$, with the property that
\begin{equation}
J^2 = \epsilon,\quad JD = \epsilon' DJ, \text{ and, if } (A,\HH,D) \text{ is even, }
J\chi = \epsilon'' \chi J. 
\end{equation}
Given $n$, the possibilities for arrays of numbers $\epsilon, \epsilon', \epsilon''\in  \{\pm1\}$ are given by the tables in section \ref{ko-dim}.
Moreover, the action of $A$ satisfies the commutation rule
\begin{equation}\label{zero-order}
[a, Jb^* J^{-1}] = 0, \quad\forall a, b\in A
\end{equation}
and the operator $D$ satisfies the order one condition
\begin{equation}
[[D, a], Jb^* J^{-1}] = 0, \quad\forall a, b\in A.
\end{equation}
A spectral triple $(A,\HH,D)$ endowed with a real structure $J$ is called a real spectral triple.
\end{defn}

\begin{rmks}~

\begin{itemize}
 \item usually we will omit to indicate the $*$-representation map $\rho\colon A\to\BB(\HH)$ for simplicity;
 \item we recall that an antiunitary operator $J$ is antilinear, bijective and $(Ju|Jv)=(v|u)$;
 \item the map $b\mapsto J b^* J^{-1}$ is a representation of the opposite algebra $A^\circ$ on $\BB(\HH)$;
 \item equation (\ref{zero-order}) which says that $\text{Ad}_J$ sends $A$ to its commutant is sometimes called the ``zero order condition'';
\item note that putting:
\begin{equation}
 \vec{\epsilon}_\pm (n):=
\left(\epsilon_\pm (n),  \epsilon'_\pm (n),  \epsilon''_\pm (n) \right),
\end{equation}
(where $n\in\IZ_8$) we have the relation:
\begin{equation}
 \vec{\epsilon}_-(n)=-\vec{\epsilon}_+(n+2)\, .
\end{equation}
\end{itemize}
\end{rmks}

\section{Product of real spectral triples}
Following what happens in the commutative case, we shall produce a real spectral triple of dimension $n_1+n_2$ out of two triples of dimensions $n_1$ and $n_2$ respectively. The new algebra is the tensor product algebra $A:=A_1\ts A_2$, where $\ts$ is the algebraic tensor product\footnote{Over $\IR$ if at least one of the two algebras is real, over $\IC$ if both algebras are complex.} and the involution is defined component-wise: $(a\ts b)^*:=a^*\ts b^*$. It turns out that the other ingredients of the resulting spectral triple depend on the parity of the two given triples.

\subsection{Even-even case}\label{sec:even-even}
As the Hilbert space carrying the $*$-representation of $A$ we take the Hilbert tensor product $\HH_1\ts\HH_2$ and as the $*$-representation we take the tensor product representation:
\begin{equation}
\rho_1\ts\rho_2\colon A_1\ts A_2\to \BB(\HH_1\ts\HH_2)\, .
\end{equation}
The representation $\rho_1\ts\rho_2$ is faithful whenever $\rho_1$ and $\rho_2$ are. The grading operator is given by $\chi:=\chi_1\ts \chi_2$ (it is easy to check that it is unitary, squares to $\id_{\HH_1\ts\HH_2}$ and commutes with every element of the product algebra $A$). As for the Dirac operator, using $\chi_1$ or $\chi_2$ we take the following operators
\footnote{The simplest choice $D=D_1\ts\id_{\mathcal{H}_2} + \id_{\mathcal{H}_1}\ts D_2$ has non-compact resolvent in general (e.g. $\ker D$ is infinite dimensional if $D_2=-D_1$ with $\HH_1=\HH_2$ infinite dimensional).}:
\begin{equation}
 \begin{aligned}
  \DD&:=D_1\ts \id_{\mathcal{H}_2} + \chi_1\ts D_2\;,\\
  \widetilde{\DD}&:=D_1\ts \chi_2 + \id_{\mathcal{H}_1}\ts D_2\, ,
 \end{aligned}
\end{equation}
both defined on the dense domain $\dom D_1\ts\dom D_2$. They are unitarily equivalent:
\begin{equation}
\widetilde{\DD}=U\DD U^\dag\, ,
\end{equation}
where (see \cite{Vanhecke})
\begin{equation}
U=\frac{1}{2}(\id_{\mathcal{H}_1}\ts\id_{\mathcal{H}_2} + \chi_1\ts\id_{\mathcal{H}_2} + \id_{\mathcal{H}_1}\ts\chi_2 - \chi_1\ts\chi_2)\, .
\end{equation}
We now show that $\DD$ is essentially self-adjoint. This is immediate if one Hilbert space is finite dimensional, so we assume both Hilbert spaces to be infinite dimensional.

From the general theory of (linear unbounded) operators on a complex separable Hilbert space we know that each $D_i$ has pure point spectrum consisting of countably many real eigenvalues, each with finite multiplicity, and the only limit point of their absolute values is $+\infty$. Let $\{v_{\lambda, m_\lambda},\;v_{j_\pm}\mid \lambda\in\sigma(D_1)/\{0\},\;m_\lambda=1,\dots, M_\lambda,\;j_\pm=1,\dots,K_\pm\in\IN\}$ be an orthonormal basis of eigenvectors of $D_1$, where $\sigma(D_1)$ is the spectrum of $D_1$, $M_\lambda$ is the multiplicity of $\lambda\in\sigma(D_1)/\{0\}$, and finally $\{v_{j_\pm}\mid j_\pm=1,\dots K_\pm\}$ is a basis of $\ker D_1$ consisting of eigenvectors of $\chi_1$ such that $\chi_1 v_{j_\pm}=\pm v_{j_\pm}$. Note that since $D_1$ and $\chi_1$ anticommute, and $\chi_1$ is unitary, then $\sigma(D_1)$ is symmetric and $M_\lambda=M_{-\lambda}$. Let $\{w_{\mu,n_\mu}\mid \mu\in\sigma(D_2),\;n_\mu=1,\dots,N_\mu\in\IN\}$ be an orthonormal basis of eigenvectors of $D_2$. Finally, let us consider the following vectors in $\HH_1\ts\HH_2$:
\begin{equation}
\begin{aligned}
u_{\lambda,m_\lambda,\mu,n_\mu}^+&:=\cos\theta_{\lambda\mu} (v_{\lambda,m_\lambda}\ts w_{\mu,n_\mu}) + \sin\theta_{\lambda\mu} (\chi_1 v_{\lambda,m_\lambda}\ts w_{\mu,n_\mu})\ ,\\
u_{\lambda,m_\lambda,\mu,n_\mu}^-&:=-\sin\theta_{\lambda\mu} (v_{\lambda,m_\lambda}\ts w_{\mu,n_\mu}) + \cos\theta_{\lambda\mu} (\chi_1 v_{\lambda,m_\lambda}\ts w_{\mu,n_\mu})\ ,\\
u_{j_\pm,\mu,n_\mu}&:=v_{j_\pm}\ts w_{\mu,n_\mu}\ ,
\end{aligned}
\end{equation}
where $\lambda\in\sigma(D_1)\cap\IR_{>0}$, $\mu\in\sigma(D_2)$, $\theta_{\lambda\mu}:=\frac{1}{2}\arctan\frac{\mu}{\lambda}\in(-\pi/4,\pi/4)$ and $j_\pm=1,\dots K_\pm$. Then the set
\begin{equation}\label{basis}
\{u_{\lambda,m_\lambda,\mu,n_\mu}^+,\;u_{\lambda,m_\lambda,\mu,n_\mu}^-,\;u_{j_\pm,\mu,n_\mu}\mid \lambda\in\sigma(D_1)\cap\IR_{>0},\;\mu\in\sigma(D_2),\;j_\pm=1,\dots, K_\pm\}
\end{equation}
is an orthonormal basis of eigenvectors of $\DD$, with corresponding eigenvalues given by:
\begin{equation}
\begin{aligned}
\DD (u_{\lambda,m_\lambda,\mu,n_\mu}^\pm)&=\pm\sqrt{\lambda^2+
\mu^2}\ u_{\lambda,m_\lambda,\mu,n_\mu}^\pm\, ,\\
\DD (u_{j_\pm,\mu,n_\mu}) &= \pm\mu\ u_{j_\pm,\mu,n_\mu}\, .
\end{aligned}
\end{equation}

It can be easily seen that $\ker \DD = \ker D_1\ts \ker D_2$. From the existence of a basis of eigenvectors for $\DD$ we can promptly conclude that $D:=\overline{\DD}$ (the closure of $\DD$) is self-adjoint. From the analysis above it is clear that $D$ is a self-adjoint operator with pure point spectrum consisting of countably many eigenvalues, each with finite multiplicity, and the only limit point of their absolute values is $+\infty$. By the general theory, we conclude that $D$ has compact resolvent.

By unitary equivalence, the basis for $\DD$ gives a basis of eigenvectors of $\widetilde{\DD}$ with the same eigenvalues and multiplicities as those of $\DD$, so the analysis above goes through and we conclude that $\widetilde{D}:=\overline{\widetilde{\DD}}$ is also self-adjoint with compact resolvent.

It is easy to check that $[D,a]$ extends to a bounded operator for each $a\in A_1\ts A_2$ (using the condition $[\chi,a]=0$). Analogously for $\widetilde{D}$.

From $J_1$ and $J_2$ we can construct $J=J_1\ts J_2$, which is easily seen to be antiunitary on $\HH_1\ts\HH_2$. Moreover we have
\begin{equation}
\begin{aligned}
~[a_1\ts a_2, J(b_1\ts b_2)^*J^{-1}]&=[a_1\ts a_2, (J_1\ts J_2)(b_1^*\ts b_2^*)(J_1^{-1}\ts J_2^{-1})]\\
&=[a_1,J_1b_1^*J_1^{-1}]\ts a_2J_2b_2^*J_2^{-1}+\\
&\quad+J_1b_1^*J_1^{-1}a_1\ts[a_2,J_2b_2^*J_2^{-1}]\\
&=0\, .
\end{aligned}
\end{equation}

Labeling the $\epsilon$-triples with $n_+$ or $n_-$ according to the KO-dimension and the $J$ involved in the product, we get the following tables for the KO-dimension and reality structure of the resulting triple (we distinguish the two cases for the total Dirac operator, $D$ or $\widetilde{D}$):

\begin{table}[htbp]\begin{center}\caption{$D$}\label{table:evenevenD}
\begin{tabular}{|c||c|c|c|c||c|c|c|c|}
 \hline
 \backslashbox{1}{2} & $0_+$ & $2_+$ & $4_+$ & $6_+$ & $0_-$ & $2_-$ & $4_-$ & $6_-$\\
 \hline
 \hline
 $0_+$ & $0_+$ & $2_+$ & $4_+$ & $6_+$ & & & & \\
 \hline
 $2_+$ & & & & & $2_+$ & $4_+$ & $6_+$ & $0_+$\\
 \hline
 $4_+$ & $4_+$ & $6_+$ & $0_+$ & $2_+$ & & & & \\
 \hline
 $6_+$ & & & & & $6_+$ & $0_+$ & $2_+$ & $4_+$\\
 \hline
 \hline
 $0_-$ & & & & & $0_-$ & $2_-$ & $4_-$ & $6_-$\\
 \hline
 $2_-$ & $2_-$ & $4_-$ & $6_-$ & $0_-$ & & & & \\
 \hline
 $4_-$ & & & & & $4_-$ & $6_-$ & $0_-$ & $2_-$\\
 \hline
 $6_-$ & $6_-$ & $0_-$ & $2_-$ & $4_-$ & & & & \\
 \hline
\end{tabular}
\end{center}
\end{table}

\begin{table}[htbp]\begin{center}\caption{$\widetilde{D}$}
\begin{tabular}{|c||c|c|c|c||c|c|c|c|}
 \hline
 \backslashbox{1}{2} & $0_+$ & $2_+$ & $4_+$ & $6_+$ & $0_-$ & $2_-$ & $4_-$ & $6_-$\\
 \hline
 \hline
 $0_+$ & $0_+$ & & $4_+$ & & & $2_-$ & & $6_-$\\
 \hline
 $2_+$ & $2_+$ & & $6_+$ & & & $4_-$ & & $0_-$\\
 \hline
 $4_+$ & $4_+$ & & $0_+$ & & & $6_-$ & & $2_-$\\
 \hline
 $6_+$ & $6_+$ & & $2_+$ & & & $0_-$ & & $4_-$\\
 \hline
 \hline
 $0_-$ & & $2_+$ & & $6_+$ & $0_-$ & & $4_-$ & \\
 \hline
 $2_-$ & & $4_+$ & & $0_+$ & $2_-$ & & $6_-$ & \\
 \hline
 $4_-$ & & $6_+$ & & $2_+$ & $4_-$ & & $0_-$ & \\
 \hline
 $6_-$ & & $0_+$ & & $4_+$ & $6_-$ & & $2_-$ & \\
 \hline
\end{tabular}
\end{center}
\end{table}

\begin{rmk}
 The two top blocks in table \ref{table:evenevenD} correspond to the even-even cases covered by Vanhecke's paper \cite{Vanhecke}.
\end{rmk}

\subsection{Even-odd case}
The Hilbert space $\HH$, the $*$-representation of $A$ on $\BB(\HH)$ and the reality structure $J$ are the same as in the even-even case. Now we have only one nontrivial grading operator though: we then choose $D$ or $\widetilde{D}$ from the previous construction, according to whether the even triple is the first one or the second one, respectively. The basis of eigenvectors is again given by (\ref{basis}) (or by the analogous construction using $\chi_2$ instead of $\chi_1$). The argument for proving self-adjointness and compactness of resolvent goes through exactly as before.
For the KO-dimension and reality structure of the resulting triple we obtain the following tables:

\begin{table}[htbp]\begin{center}\caption{$D$}\label{table:evenoddD}
\begin{tabular}{|c||c|c|c|c|}
 \hline
 \backslashbox{1}{2} & $1_-$ & $3_+$ & $5_-$ & $7_+$\\
 \hline
 \hline
 $0_+$ &  & $3_+$ &  & $7_+$\\
 \hline
 $2_+$ & $3_+$ &  & $7_+$ & \\
 \hline
 $4_+$ &  & $7_+$ &  & $3_+$\\
 \hline
 $6_+$ & $7_+$ &  & $3_+$ & \\
 \hline
 \hline
 $0_-$ & $1_-$ &  & $5_-$ & \\
 \hline
 $2_-$ &  & $5_-$ &  & $1_-$\\
 \hline
 $4_-$ & $5_-$ &  & $1_-$ & \\
 \hline
 $6_-$ &  & $1_-$ &  & $5_-$\\
 \hline
\end{tabular}
\end{center}
\end{table}

\begin{table}[htbp]\begin{center}\caption{$\widetilde{D}$}
\begin{tabular}{|c||c|c|c|c||c|c|c|c|}
 \hline
 \backslashbox{1}{2} & $0_+$ & $2_+$ & $4_+$ & $6_+$ & $0_-$ & $2_-$ & $4_-$ & $6_-$\\
 \hline
 \hline
 $1_-$ &  & $3_+$ &  & $7_+$ & $1_-$ &  & $5_-$ & \\
 \hline
 $3_+$ & $3_+$ & & $7_+$ & & & $5_-$ & & $1_-$\\
 \hline
 $5_-$ &  & $7_+$ &  & $3_+$ & $5_-$ &  & $1_-$ & \\
 \hline
 $7_+$ & $7_+$ & & $3_+$ & & & $1_-$ & & $5_-$\\
 \hline
\end{tabular}
\end{center}
\end{table}

\begin{rmk}
 Table \ref{table:evenoddD} corresponds to the even-odd cases covered by Vanhecke's paper \cite{Vanhecke}.
\end{rmk}

\subsection{Odd-odd case}
In this case we have no nontrivial grading operator available. In order to overcome this, motivated by the commutative situation, we consider the following construction:
\begin{equation}
 \begin{aligned}
  A&:=A_1\ts A_2,\\
  \mathcal{H}&:=(\mathcal{H}_1\ts \mathcal{H}_2)\ts \complex^2,\\
  \DD&:=D_1\ts \id_{\mathcal{H}_2}\ts \sigma_1 + \id_{\mathcal{H}_1}\ts D_2\ts \sigma_2,\\
  J^\pm&:=J_1\ts J_2\ts M^\pm K,\\
  \chi&:=\id_{\mathcal{H}_1}\ts \id_{\mathcal{H}_2}\ts \sigma_3,
 \end{aligned}
\end{equation}
where the $\sigma$s are the Pauli matrices, $M^\pm$ are two complex matrices specified by the table below and $K$ is the complex conjugation operator defined for the canonical basis of $\complex^2$ (i.e., if $(e_1, e_2)$ is the canonical basis, we have $K(\lambda e_i)=\overline{\lambda}e_i$ for every $\lambda\in\complex$). The representation is understood to be trivial on the $\complex^2$ factor, i.e. $\rho(a_1\ts a_2)=\rho_1(a_1)\ts\rho_2(a_2)\ts \id_{\complex^2}$.

\begin{table}[htbp]\begin{center}\caption{Odd-odd case}\label{table:oddodd}
\begin{tabular}{|c||c|c|c|c|}
 \hline
 \backslashbox{1}{2} & $1_-$ & $3_+$ & $5_-$ & $7_+$\\
 \hline
 \hline
 $1_-$ & $\sigma_2, \sigma_1$ & $\sigma_3, \sigma_0$ & $\sigma_2, \sigma_1$ & $\sigma_3, \sigma_0$\\
 \hline
 $3_+$ & $\sigma_0, \sigma_3$ & $\sigma_1, \sigma_2$ & $\sigma_0, \sigma_3$ & $\sigma_1, \sigma_2$\\
 \hline
 $5_-$ & $\sigma_2, \sigma_1$ & $\sigma_3, \sigma_0$ & $\sigma_2, \sigma_1$ & $\sigma_3, \sigma_0$\\
 \hline
 $7_+$ & $\sigma_0, \sigma_3$ & $\sigma_1, \sigma_2$ & $\sigma_0, \sigma_3$ & $\sigma_1, \sigma_2$\\
 \hline
\end{tabular}
\end{center}
\end{table}

\begin{rmks}
The entries in table \ref{table:oddodd} stand for the pair $M^+$, $M^-$.
For convenience, the identity matrix is called $\sigma_0$. Note that this construction still works under any permutation of the Pauli matrices (e.g., one can take $\DD:=D_1\ts \id_{\mathcal{H}_2}\ts \sigma_1 + \id_{\mathcal{H}_1}\ts D_2\ts \sigma_3$ and $\chi:=\id_{\mathcal{H}_1}\ts \id_{\mathcal{H}_2}\ts \sigma_2$).
The table obtained considering only the first element in each entry (i.e. $M^+$) corresponds to the odd-odd cases covered by Sitarz's notes \cite{Sitarz}.
\end{rmks}

Calling $n_1=2m_1+1$ and $n_2=2m_2+1$ the dimensions of the two triples involved, we have:
\begin{equation}
\begin{aligned}
 M^+(n_1,n_2)&=\sigma_j\;,\quad j={\textstyle\frac{1}{2}}\left(5+(-1)^{m_2+1}\right)+2m_1\mod 4\, ,\\
 M^-(n_1,n_2)&=\sigma_k\;,\quad k={\textstyle\frac{1}{2}}\left(1+(-1)^{m_2}\right)+2m_1\mod 4\, .
\end{aligned}
 \end{equation}

Self-adjointness of $D:=\overline{\DD}$ and compactness of its resolvent can be proven by the same argument of section \ref{sec:even-even} with suitable changes. In particular the eigenvectors $u^\pm$ are given by:
\begin{equation}
\begin{aligned}
u_{\lambda,m_\lambda,\mu,n_\mu}^+&:=\frac{1}{\sqrt{2}}\cos\theta_{\lambda\mu} v_{\lambda,m_\lambda}\ts w_{\mu,n_\mu}\ts\binom{1}{1} + \frac{1}{\sqrt{2}}\sin\theta_{\lambda\mu} v_{\lambda,m_\lambda}\ts w_{\mu,n_\mu}\ts\binom{-i}{i}\ ,\\
u_{\lambda,m_\lambda,\mu,n_\mu}^-&:=-\frac{1}{\sqrt{2}}\sin\theta_{\lambda\mu} v_{\lambda,m_\lambda}\ts w_{\mu,n_\mu}\ts\binom{1}{1} + \frac{1}{\sqrt{2}}\cos\theta_{\lambda\mu} v_{\lambda,m_\lambda}\ts w_{\mu,n_\mu}\ts\binom{-i}{i}\ .
\end{aligned}
\end{equation}

\section{Further properties and their preservation under products}\label{7axioms}

\begin{itemize}
  \item \textbf{Dimension} (``condition 1'' on p. 481 in \cite{G-BVF}): Assume $(A_i, \mathcal{H}_i, D_i, J_i, (\chi_i))_{i=1,2}$ are two real spectral triples of dimensions $n_1$ and $n_2$ respectively. Independently of the parities of the triples, the eigenvalues of $D^2$ are given by the sum of the eigenvalues of $D_1^2$ and $D_2^2$, and this implies that the dimension of the product triple is $n_1+n_2$ (see p. 486 in \cite{G-BVF} for details).
 \item \textbf{Regularity} (``condition 2'' on p. 482 in \cite{G-BVF}):
For even spectral triples the result that the product of two regular triples is regular is contained in \cite{Otogo}, which uses the existence of an algebra of generalized differential operators; this works also when at least one of the two triples is even and for the odd-odd case the argument still can be carried over.

 \item \textbf{Reality} (``condition 4'' on p. 483 in \cite{G-BVF}): This condition is automatically preserved, by construction of the product triple.
 \item \textbf{First order} (``condition 5'' on p. 484 in \cite{G-BVF}): Assume $(A_i, \mathcal{H}_i, D_i, J_i, \chi_i)_{i=1,2}$ are two real spectral triples of even dimension satisfying this property. Then the product $(A, \mathcal{H}, D, J, \chi)$ satisfies it as well: indeed, taking $a,b\in A$, where $a=a_1\ts a_2$, $b=b_1\ts b_2$ we compute:
\begin{equation}
 \begin{aligned}
  \left[\left[ D, a\right], b^\circ\right]&=\left[\left[ D_1\ts \id_{\mathcal{H}_2}+\chi_1\ts D_2, a\right], b^\circ\right]\\
&=\left[\left[ D_1\ts\id_{\mathcal{H}_2}, a\right] + \left[ \chi_1\ts D_2, a \right], b^\circ\right]\\
&=\left[\left[ D_1, a_1\right]\ts a_2 + \chi_1 a_1\ts\left[D_2, a_2 \right], b^\circ\right]\\
&=\left[\left[ D_1, a_1\right]\ts a_2, b^\circ\right] + \left[\chi_1 a_1\ts\left[D_2, a_2 \right], b^\circ\right]\\
&=\left[\left[ D_1, a_1\right],b_1^\circ\right]\ts a_2 b_2^\circ + \chi_1 a_1 b_1^\circ\ts\left[\left[D_2, a_2\right], b_2^\circ\right]\\
&=0,
 \end{aligned}
\end{equation}
where $b^\circ:=Jb^*J^{-1}$, where we have used the facts that the representations of $A_i$ and $A_i^\circ$ commute (this is part of the content of condition 5) and also that $\chi_i$ commutes with the representation of $A_i$. Analogously one can prove that the representations of $A$ and $A^\circ$ commute, i.e. $\left[a, b^\circ\right]=0$. The same computation also applies for $\widetilde{D}$, and this concludes the proof for the even-even case. For the even-odd case the computations are analogous. For the odd-odd case we compute:
\begin{equation}
 \begin{aligned}
  \left[\left[ D, a\right], b^\circ\right]&=\left[\left[ D_1\ts\id_{\mathcal{H}_2}\ts\sigma_1+\id_{\mathcal{H}_1}\ts D_2\ts\sigma_2, a\right], b^\circ\right]\\
&=\left[\left[D_1\ts\id_{\mathcal{H}_2}\ts\sigma_1,a_1\ts a_2\ts \id_{\complex^2}\right], b_1^\circ\ts b_2^\circ\ts\id_{\complex^2}\right]+\\
&\quad+\left[\left[\id_{\mathcal{H}_1}\ts D_2\ts\sigma_2,a_1\ts a_2\ts \id_{\complex^2}\right], b_1^\circ\ts b_2^\circ\ts\id_{\complex^2}\right]\\
&=\left[\left[D_1,a_1\right]\ts a_2\ts\sigma_1, b_1^\circ\ts b_2^\circ\ts\id_{\complex^2}\right]+\\
&\quad+\left[a_1\ts \left[D_2,a_2\right]\ts\sigma_2, b_1^\circ\ts b_2^\circ\ts\id_{\complex^2}\right]\\
&=\left[\left[D_1,a_1\right],b_1^\circ\right]\ts a_2 b_2^\circ\ts\sigma_1+\\
&\quad+a_1 b_1^\circ\ts\left[\left[D_2,a_2\right],b_2^\circ\right]\ts\sigma_2\\
&=0,
 \end{aligned}
\end{equation}
where we used the fact that the reps of $A_i$ and $A_i^\circ$ commute. As before, checking that the reps of $A$ and $A^\circ$ commute is entirely analogous.

\item \textbf{Orientation} (``condition 6'' on p. 484 in \cite{G-BVF}): Given two Hochschild cycles $a_i\in Z_{n_i}(A_i, A_i\ts A_i^\circ)$, $i=1,2$, where $n_i$ is the dimension of the algebra $A_i$ according to ``condition 1'', one can construct the Hochschild cycle $a\in Z_n(A, A\ts A^\circ)$ (where $n=n_1+n_2$ and $A=A_1\ts A_2$) using the shuffle product (see \cite{Loday}, section 4.2). Let us provide some details for the construction, following \cite{Loday}. First define\footnote{For the sake of generality, here we can take the first coefficient $a_0$ to be in some module over $A$; we will be interested in the case where this module is $A\ts A^\circ$.} the shuffle product $\times \colon C_{n_1}(A_1)\ts C_{n_2}(A_2)\to C_{n_1+n_2}(A_1\ts A_2)$ of two chains as follows:
\begin{equation}
\begin{aligned}
 &(a^1_0, a^1_1, \dots, a^1_p)\times(a^2_0, a^2_1, \dots, a^2_q):=\\
&\sum_\sigma (-1)^\sigma \sigma\cdot(a^1_0\ts a^2_0, a^1_1\ts 1, \dots, a^1_p\ts 1,1\ts a^2_1, \dots, 1\ts a^2_q),
\end{aligned}
\end{equation}
where
\begin{equation}
 \sigma\cdot(a_0,a_1\dots, a_n):=(a_0, a_{\sigma^{-1}(1)}, \dots, a_{\sigma^{-1}(n)})
\end{equation}
and the sum is over all $(p,q)$-shuffles, i.e. permutations of $\{1,\dots, p+q\}$ preserving the order of $\{1,\dots, p\}$ and $\{p+1,\dots, p+q\}$ separately. Then for the Hochschild boundary map $\partial$ the following formula holds:
\begin{equation}
 \partial (x\times y)=\partial(x)\times y + (-1)^{|x|}x\times \partial(y)\, .
\end{equation}
From this formula it follows at once that the shuffle product of two Hochschild cycles is again a Hochschild cycle. The orientation condition for a triple now states that there is a Hochschild cycle $c$ satisfying the following formula:
\begin{equation}\label{eq:orientation}
\pi_D(c)=\chi,
\end{equation}
where the map $\pi_D$ is defined as:
\begin{equation}
\pi_{D}(a_0\ts a_1\ts\cdots a_p)=\tau_J(a_0)[D,\rho(a_1)]\cdots [D,\rho(a_p)],
\end{equation}
where $a_0\in A\ts A^\circ$ and $\tau_J(a\ts b):=\rho(a)J\rho(b^*)J^{-1}$.
Given two real spectral triples of dimensions $n_1$ and $n_2$ respectively, we claim that if $c_j$ ($j=1,2$) are Hochschild cycles satisfying $\pi_{D_j}(c_j)=\chi_j$ then the analogous cycle on the product triple is given by
\begin{equation}\label{eq:product_c}
c:=\textstyle{\frac{1}{r}} c_1\times c_2,
\end{equation}
where
\begin{equation}\begin{aligned}
r&:=\begin{cases}
\nu_{n_1+n_2}\ , & \text{when }n_1n_2\text{ is even}\\
i\nu_{n_1+n_2}\ , & \text{when } n_1n_2\text{ is odd}\ ,
\end{cases}\\
\nu_n&:=\frac{1}{2}(n-1)n\ ,
\end{aligned}
\end{equation}
where $\nu(n):=\frac{1}{2}(n-1)n$. In order to check formula (\ref{eq:orientation}) on the product triple with $c$ given by equation (\ref{eq:product_c}) we distinguish three cases depending on the parities involved:
\begin{description}
\item[Even-even.] A simple computation shows that
\begin{equation}
\begin{aligned}
~[D,\rho_1(a)\ts \id_{\mathcal{H}_2}]&=[D_1,\rho_1(a)]\ts \id_{\mathcal{H}_2}\\
[D,\id_{\mathcal{H}_1}\ts \rho_2(b)]&=\chi_1\ts [D_2,\rho_2(b)]
\end{aligned}
\end{equation}
from which it follows that
\begin{equation}
\begin{aligned}
&\pi_D(\sigma\cdot(a_0\ts b_0, a_1\ts 1, \dots, a_{n_1}\ts 1,1\ts b_1, \dots, 1\ts b_{n_2}))=\\
=&\Pi\sigma\cdot\left(\tau_{J_1}(a_0), [D_1, \rho_1(a_1)], \dots, [D_1, \rho_1(a_{n_1})], \chi_1, \dots, \chi_1\right)\ts\\
&\ts \Pi(\tau_{J_2}(b_0),[D_2, \rho_2(b_1)], \dots, [D_2, \rho_2(b_{n_2})]),
\end{aligned}
\end{equation}
where $\Pi$ means algebra product of all the elements in the ordered list. Since $\chi_1$ anti-commutes with $D_1$ and commutes with $\rho_1(a)$ for each $a\in A_1$, rearranging all the $n_2$ operators $\chi_1$ side by side produces a $(-1)^\sigma$ sign which cancels the same sign from the shuffle product; moreover, since $n_2$ is even, we have ${\chi_1}^{n_2}=\id_{\mathcal{H}_1}$; therefore we are left with a sum of $\frac{1}{2}(n_1+n_2-1)(n_1+n_2)$ identical terms:
\begin{equation}
\begin{aligned}
\pi_D(c_1\times c_2)&=\nu_{n_1+n_2}\ \pi_{D_1}(c_1)\ts \pi_{D_2}(c_2)\\
&=\nu_{n_1+n_2}\ \chi_1\ts \chi_2\, .
\end{aligned}
\end{equation}
The same reasoning applies to $\widetilde{D}$ with obvious modifications.
\item[Even-odd.] The previous argument carries over unaltered, but this time we have ${\chi_1}^{n_2}=\chi_1$ since $n_2$ is odd. Then we get
\begin{equation}
\begin{aligned}
\pi_D(c_1\times c_2)&=\nu_{n_1+n_2}\ \chi_1\pi_{D_1}(c_1)\ts \pi_{D_2}(c_2)\\
&=\nu_{n_1+n_2}\ {\chi_1}^2\ts \chi_2\\
&=\nu_{n_1+n_2}\ \id_{\mathcal{H}_1}\ts (\pm\id_{\mathcal{H}_2})\\
&=\pm\nu_{n_1+n_2}\ \id_{\mathcal{H}_1}\ts \id_{\mathcal{H}_2}\ ,
\end{aligned}
\end{equation}
as expected. The same reasoning applies to the case where $n_1$ is odd and $n_2$ is even, with obvious modifications.
\item[Odd-odd.] In this case a simple computation shows that
\begin{equation}
\begin{aligned}
~[D,\rho_1(a)\ts \id_{\mathcal{H}_2}\ts \id_{\complex^2}]&=[D_1,\rho_1(a)]\ts \id_{\mathcal{H}_2}\ts \sigma_1\\
[D,\id_{\mathcal{H}_1}\ts \rho_2(b)\ts \id_{\complex^2}]&=\id_{\mathcal{H}_1}\ts [D_2,\rho_2(b)]\ts \sigma_2
\end{aligned}
\end{equation}
from which it follows that
\begin{equation}
\begin{aligned}
&\pi_D(\sigma\cdot(a_0\ts b_0, a_1\ts 1, \dots, a_{n_1}\ts 1,1\ts b_1, \dots, 1\ts b_{n_2}))=\\
=&\tau_{J_1}(a_0)[D_1, \rho_1(a_1)]\cdots[D_1, \rho_1(a_{n_1})]\ts\\
&\ts\tau_{J_2}(b_0)[D_2, \rho_2(b_1)]\cdots [D_2, \rho_2(b_{n_2})])\ts\\
&\ts\Pi\sigma\cdot(1,\underbrace{\sigma_1,\dots, \sigma_1}_{n_1\text{ times}},\underbrace{\sigma_2,\dots, \sigma_2}_{n_2\text{ times}})
\end{aligned}
\end{equation}
\end{description}
\end{itemize}
Since $\sigma_1\sigma_2=-\sigma_2\sigma_1$ we can rearrange the $\sigma_i$s with all $\sigma_1$s on the left and all $\sigma_2$s on the right, producing a $(-1)^\sigma$ sign which cancels the same sign from the shuffle product; moreover since $n_1$ and $n_2$ are both odd we get $\sigma_1^{n_1}\sigma_2^{n_2}=\sigma_1\sigma_2 = i \sigma_3$, so we end up with
\begin{equation}
\begin{aligned}
\pi_{D}(c_1\times c_2) &= \nu_{n_1+n_2}\ \pi_{D_1}(c_1)\ts \pi_{D_2}(c_2)\ts i\sigma_3\\
&= i\nu_{n_1+n_2}\ \id_{\mathcal{H}_1}\ts \id_{\mathcal{H}_2}\ts \sigma_3\, .
\end{aligned}
\end{equation}

\begin{rmk}
 The orientation axiom is consistent with the observation made in section \ref{Dirac_op} for the classical setting, namely that changing the representation $\Gamma_{(2m+1,+)}$ to $\Gamma_{(2m+1,-)}$ is equivalent to changing the orientation of the manifold. In the noncommutative setting, this translates into changing the sign of the Hochschild cycle $c$ in (\ref{eq:orientation}).
\end{rmk}

\section{Final comments}

In this paper we are concerned with unital spectral triples but the canonical Dirac operator on $\IR^n$ (cf. section \ref{Dirac_op}) is not of that type. The definition of a nonunital spectral triple is slightly different, as well as the additional axioms for it.
In order to remain in the realm of unital spectral triples, as a commutative compact case study we should take rather the flat torus with the trivial spin structure.
This however does not change the form (\ref{Dop}) of the canonical Dirac operator, but just supplements it with periodic boundary conditions.

It is worth mentioning that in our setup the metric and KO dimensions need not be equal modulo 8. This is the case in some of the recent
examples of spectral triples \cite{DS}, \cite{C07, C08}, see also \cite{Cacic}.

\section*{Acknowledgments}
We wish to thank Jyotishman Bhowmick and Francesco D'Andrea for useful discussions, and Andrzej Sitarz for informing us about the unpublished notes \cite{Sitarz}.

%\vskip 2cm

\end{document}